\documentclass[10pt,sigconf]{acmart}

\usepackage[english]{babel}
\usepackage[inline]{aplcomments}
\usepackage{xspace} 
\usepackage{enumitem}
\usepackage{listings}
\usepackage{xcolor}
\usepackage{dsfont}
\usepackage{amsmath}
\usepackage{stmaryrd}
\usepackage{amsfonts}
\usepackage{mathtools}
\usepackage{algorithm}
\usepackage{hyperref}
\usepackage{adjustbox}
\usepackage{url}
\usepackage{graphicx}

\definecolor{lightgray}{rgb}{.9,.9,.9}
\definecolor{darkgray}{rgb}{.4,.4,.4}
\definecolor{purple}{rgb}{0.65, 0.12, 0.82}

\lstdefinelanguage{solidity}{
  keywords={contract, new, true, false, catch, function, return, null, catch, switch, var, if, in, while, do, else, case, break},
  keywordstyle=\color{blue}\bfseries,
  ndkeywords={class, export, boolean, throw, implements, import, this, bool, public, constant, returns, uint, uint256, address},
  ndkeywordstyle=\color{darkgray}\bfseries,
  identifierstyle=\color{black},
  sensitive=false,
  comment=[l]{//},
  morecomment=[s]{/*}{*/},
  basicstyle=\footnotesize\ttfamily,
  commentstyle=\color{purple}\ttfamily,
  stringstyle=\color{red}\ttfamily,
  morestring=[b]',
  morestring=[b]"
}

\newcommand{\figlabel}[1]{\label{fig:#1}}
\newcommand{\seclabel}[1]{\label{sec:#1}}

\newcommand{\secref}[1]{Sec.~\ref{sec:#1}}  
\newcommand{\figref}[1]{Fig.~\ref{fig:#1}}     

\newcommenter{shumo}{1.0,0.4,1.0} 
\newcommenter{sophia}{0.4,0.4,1.0} 
\newcommenter{xi}{1.0,0.4,0.4} 

\renewcommand\footnotetextcopyrightpermission[1]{} 
\setcopyright{none}

\acmDOI{}

\acmISBN{}

\acmPrice{}

\begin{document}
\title{The Curses of Blockchain Decentralization}


\author{Shumo Chu, Sophia Wang}
\authornote{This work doesn't represent the opinion of the authors' employers. Epichain is a blockchain initiative started by the authors (\url{http://epichain.io}).}
\affiliation{
\institution{University of Washington and Epichain.io}
}

\email{shumo.chu@acm.org, wangxiao@cs.washington.edu}

\renewcommand{\shortauthors}{Chu and Wang}
\newcommand{\heading}[1]{\vspace{4pt}\noindent\textbf{#1}}

\begin{abstract}
Decentralization, which has backed the hyper growth of many blockchains,
comes at the cost of scalability.
To understand this fundamental limitation,
this paper proposes a quantitative measure of blockchain decentralization,
and discusses its implications to various trust models and
consensus algorithms.
Further, we identify the major challenges in blockchain decentralization.
Our key findings are that true decentralization is hard to achieve 
due to the skewed mining power and that a fully decentralized
blockchain inherently limits scalability as it incurs a throughput upper bound
and prevents scaling smart contract execution.
To address these challenges,
we outline three research directions to explore the trade-offs
between decentralization and scalability.

\end{abstract}

\maketitle


\section{Introduction}

Since the invention of Bitcoin \cite{bitcoin}, 
cryptocurrencies are gaining enormous 
popularity. As of April 2018, two most popular cryptocurrencies, 
Bitcoin and Ethereum \cite{ether}, have 135.8 and 49.8 billions USD 
market capitalization, respectively \cite{coincaps}.
One key innovation pioneered by Bitcoin and adopted by many other 
cryptocurrencies is running consensus protocols with open membership
on top of cryptographic data structures called \emph{blockchains}.
As a result, trust of the currencies can be shifted from centralized control
to crypto properties together with the consensus protocol.

The idea of combining consensus protocols and 
blockchains reaches far beyond cryptocurrencies. 
This idea was extended to build secure ledgers 
for generalized transactions, namely, \emph{smart contracts}.
And according to the \emph{decentralization level} 
of blockchain providers, 
blockchain systems can be classified into public chains, 
consortium chains, and private chains.
Public chains usually use incentivized consensus protocols, 
such as Nakamoto consensus \cite{bitcoin}, that allow anyone to join.
Consortium chains only allow permissioned participants join the consensus process.
And there is only a single participant or dictator 
in consensus of private chains.

Strong decentralization enables the freedom of not trusting any 
particular blockchain providers or authorities while still ensuring the trustworthy of the whole system. 
However, this freedom is not free. 
In this paper, we aim to address the following questions. 

\heading{How to quantify decentralization?} 
We first define \emph{centralization level}, a quantitative measure 
that captures the extent of centralization of blockchains. 
This measure reflects the distributions of transactions contributed 
by blockchain providers. 
Then, we conduct case studies and compare the centralization levels 
of different blockchains. 
(\secref{decentralizaiton}). 

\heading{What are the problems with decentralization?}
We discuss blockchains in details by breaking down its system stack
into multiple layers: physical nodes, 
platform software, smart contracts, and clients. 
We report three major problems with decentralizaiton in these layers:
1) In physical layer, the assumption of decentralization of mining power does
not hold since the real-world mining power distribution is highly skewed. 
2) In platform software layer, decentralization causes 
inherent scalability problems of transaction throughput. 
We prove a low upper bound of transaction throughput of decentralized blockchains, 
which is independent of the choices of specific protocols.
3) In smart contract layer, current decentralized blockchains do fully replicated
execution and sequential programming models, which prevent scaling the smart contract execution. 
(\secref{problems}).

\heading{Research opportunities.}
These problems highlight key challenges in blockchain research. 
There are several worth exploring directions.
For example,
to democratize mining powers, new crypto hash algorithms 
that are hard for ASIC exploit significant marginal performance could be used.
In order to overcome the scalability of transaction throughput, we should explore alternative means of 
ensuring trust, for example, formal verification, verifiable computation, and secured hardware, 
rather than purely rely on decentralizaiton.  
To scale smart contract execution, a promising direction is to
co-design new programming model and runtime for smart contract that 
allows parallel execution (\secref{beyond}). 




\section{Background}
\seclabel{overview}

A blockchain is a distributed ledger of transactions that provides both
a trusted computing platform and a tamper-proof history.
To provide a trusted computing platform, a blockchain often requires
multiple system nodes to perform the same transaction and builds consensus
on the majority.
To provide a tamper-proof history, a blockchain uses chained hashes to provide
crypto proofs of serialized transactions
to defend against double-spending attacks.

\heading{Bitcoin}~\cite{bitcoin} is generally considered the first widely used blockchain.
To perform a transaction, a user requests system nodes to send Bitcoins from her address to another address, and also attaches a signature to prove its authenticity.
A system node (a.k.a., miner) makes sure that Bitcoins are spent by rightful owners and if so execute the transaction.
To commit the results of transactions to the ledger, an honest miner selects
the longest chain produced by all miners, validates existing transactions
on the chain, appends a block of transactions to the chain, and produces a hash
that wraps the new block and the hash of the previous block.
Miners are required to solve crypto-puzzles, known as Proof of Work (PoW),
to make sure that one block is built about every ten minutes globally.
To reach consensus from open pools of miners, Bitcoin economically incentivizes
miners to append to the longest chain, which is called Nakamoto consensus~\cite{bitcoin}.
Nakamoto consensus tolerates Byzantine faults as long as more than 50\% computing power
is controlled by honest system nodes.

\heading{Ethereum}~\cite{ether} is a blockchain that generalizes
transactions of cryptocurrencies into generic state
transitions, while sticking to most system components being proposed in Bitcoin.
To support generic state transitions, Ethereum proposed smart contracts
written in the Solidity programming language~\cite{solidity}.
Transaction fees (a.k.a., gas) depend on the amount of system resources
being consumed, which are calculated by Ethereum virtual machine.
Smart contracts are abstracted as accounts; running a smart contract is
similar to depositing to an account.
Similar to Bitcoin, the Ethereum system makes sure that a smart contract is correctly executed
and that state of a smart contract is tamper-proof. 
There are blockchain platforms that provides similar functionalities like Ethereum but uses different consensus protocols. For example, EOS \cite{eos} replaces Ethereum's 
Nakamoto consensus with delegated proof of stake (see more details in \secref{pos}).

\heading{Hyperledger Fabric.} Not all applications require the mining process to be open.
IBM introduced Hyperledger Fabric~\cite{hyperledger}
that uses permissioned nodes to build blockchains and that supports
smart contracts written in regular programming languages.
There are two key differences between Fabric and fully open blockchains.
One, Fabric does not include economics in its design,
meaning that the system runs without economic incentives.
Two, the permissioned Fabric does not need
Nakamoto consensus but instead uses traditional consensus algorithms (e.g.,
practical Byzantine fault tolerance).
Hyperledger Fabric is favored by closed consortiums (e.g., financial institutions).

\heading{Summary.}
Depending on the openness of system nodes,
blockchains can be classified into public, consortium, and private chains.
Bitcoin and Ethereum are public blockchains that run on P2P networks.
Public chains do not require app developers or app users to trust miners,
as long as more than 50\% of computing power is not compromised.
Hyperledger Fabric is consortium blockchains that are run by permissioned nodes.
In consortium chains, a system node is often a member of a consortium.
Private chains run on a single system node, and are thus permissioned.
Both app developers and app users have to trust the node not to compute
wrong results, not to deny a transaction, and not to tamper the history.
Therefore, the trust model that private chains provide is indifferent
from that of a cloud provider.


\section{An Analysis of Decentralization}
\seclabel{decentralizaiton}

Decentralization has been a key component of blockchains to democratize trust.
Specifically, much attention has been paid to the decentralization of system nodes.
This section formally defines decentralization as a quantitative measure,
and uses this measure to analyze a few blockchain improvements.

\subsection{Measure}

\begin{table*}[t]
\centering \footnotesize{ \fontsize{8pt}{\baselineskip}\selectfont
\begin{tabular}{llllll}
\toprule
	\textbf{Type} & \textbf{Centralization Level} & \textbf{Central Trust} & \textbf{Consensus} & \textbf{Mining} & \textbf{Examples}\\
\midrule
	Public & $N_\epsilon, \exists~c, \epsilon > 0 \to N_\epsilon>c$  & $N_{0.49}$ & Nakamoto & PoW, PoS, DPoS & Bitcoin, Ethereum, EOS\\
	Consortium & $N_\epsilon, \exists~c \to N_0<c$ & $N_{0.33}$ & PBFT & N/A & Hyperledger Fabric\\
	Private & $N_0=1$ & $1$ & N/A & N/A & N/A\\
\bottomrule
\end{tabular}
}
\caption{Centralization levels, trust models, and consensus algorithms of three kinds of blockchains.}
\label{tab:taxomony}
\vspace{-0.2in}
\end{table*}

\begin{figure}
\centering
\includegraphics[width=0.8\columnwidth]{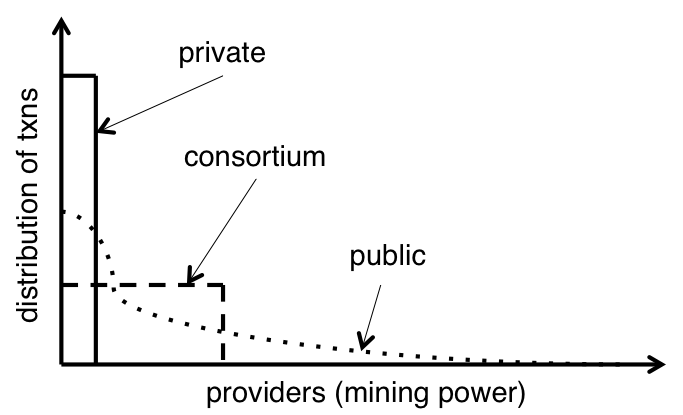}
\caption{Cumulative distributions of the number of transactions by providers (mining power).}
\figlabel{decentralization}
\vspace{-0.2in}
\end{figure}

We use \emph{centralization level} to formally define decentralization of blockchains.
A blockchain is $N_\epsilon$ centralized if the top $N$ nodes
performed more than $1-\epsilon$ fraction of transactions.
Given the same $\epsilon$, a blockchain is more centralized if it has a smaller $N$.
\figref{decentralization} shows the centralization of public, consortium, and private chains respectively.
As long as a small enough $\epsilon$ is given, a public chain's centralization
level can incur a large $N$, meaning a low level of centralization.
However, a consortium chain often incurs a small centralization level $N_0$ when $\epsilon=0$.
An extreme case of consortium chain is private chain, which is fully centralized ($N_0=1$).

Based on the definition of centralization level, we further look at
the level of central trust.
Nakamoto consensus requires 51\% of compute power, or any form of mining, to be trusted in order to tolerate Byzantine faults.
Thus, a public chain's level of central trust is $T=N_{0.49}$.
Practical Byzantine fault tolerance (PBFT) requires $(2n+1)/(3n+1) \approx 67\%$ nodes to be trusted to tolerate Byzantine faults.
Thus, a consortium chain's central trust level is $T=N_{0.33}$.
Similarly, a private chain's central trust level is $T=1$.
Here, lower trust level means more central trust.
%

%
Table~\ref{tab:taxomony} summarizes centralization levels, central trust,
and consensus algorithms of the three kinds of blockchains.
The formal definition of decentralization enables us to quantitatively
analyze decentralization and scalability of multiple blockchain improvements.

\subsection{Analysis of PoS and DPoS}
\seclabel{pos}

The process to solve a crypto-puzzle (Proof of Work) is computationally
intensive.
Thus, Proof of Stake (PoS) is proposed to save compute power.
When building a block, a miner needs to deposit stake into a contract to win the block-building chance, which is proportional to the amount of stake.
A faulty participant's deposit (stake) will be taken to penalize misbehavior.
The centralization level of a PoS blockchain depends on the distribution of stake,
instead of that of compute power which is highly skewed (\secref{problem1}).
%

Delegated Proof of Stake (DPoS) does not require every participant to directly build blocks.
Instead, participants can delegate a miner.
DPoS reduced the number of miners in consensus,
but at the same time makes a participant
harder to directly build a block,
which incurs an impact to the trust model.
As Delegated Proof of Stake reduced the number of miners to build blocks,
the centralization level would increase.
For example, EOS allows twenty-one large entities to build blocks, incurring a centralization level of $N_0=21$ which is more centralized than fully open blockchains that run PoW or PoS.

\subsection{Analysis of Sharding}

It is well known that Bitcoin and Ethereum have limited scalability,
with less than fifteen transactions per second~\cite{princetonbook, etherpeak}.
A key issue is that all the system nodes work on one blockchain.
Similar to other distributed systems, \emph{sharding}~\cite{ethersharding}, or partitioning,
would be a promising approach to improve scalability.
With sharding, the system requires a transaction to be verified only
by a small subset of nodes, and thus handles multiple transactions in parallel.
Although no implementation is finished yet, blockchain communities
consider sharding as the future of blockchains.

Let us assume that a blockchain system is evenly sharded into $k$ partitions.
If the throughput (transactions per second) of a blockchain without sharding is $t$,
that of the sharded blockchain would be $kt$.
Because only $1/k$ nodes would join the process of performing a transaction,
the centralization level with sharding would be $N_{\epsilon}/k$.
This means that the \emph{centralization-throughput product (CTP)}
stays constant given an evenly sharded blockchain.
If a blockchain is not evenly sharded, the worst case would incur a
smaller CTP, meaning either more centralized or less scalable.

\subsection{Analysis of Lightning Network}

Lightning Network~\cite{lightning} improved scalability by
offloading transactions off the blockchain (off-chain).
To do so, two clients can set up a channel on the main blockchain (on-chain)
indicating an upper bound of payments and a timeout.
It is guaranteed cryptographically that payments within the upper bound
and the timeout would be secure even if payments were taken off-chain.
As of the setup, the two clients can make payments off-chain.
When the channel times out, the two clients need another on-chain transaction
to settle the payments.
If two clients perform transactions frequently, they would benefit from
the batch processing which only incurs two on-chain transactions.
In addition to batch processing, Lightning Network enables a third party to relay payments as long as it has set up channels with the involved users.
Relaying payments would save setting up direct channels on-chain, further reducing on-chain transactions.

If the throughput of a blockchain without Lightning Network is $t$,
that with Lightning Network and no relays would be $t\alpha$ where
$\alpha$ is the compression level benefited from batch processing.
Lightning Network further improves throughput with payment relays,
which adds an additional layer of relay nodes that could withhold
transactions.
No matter what the topology of the payment graph is, Lightning Network
with $n$ clients can at most reduce the number of on-chain transactions
to $O(n)$, by using one relay node to relay all payments.
Thus, a fully connected payment graph ($O(n^2)$) would benefit most in
throughput.
At the same time, the centralization level of relay nodes would change
from $N_0=n$ to $N_0=1$.
The \emph{centralization-throughput product} in this extreme case would be $nt\alpha$,
the improvement of which comes only from batch processing.
Note that centralization of the relay nodes can only withhold transactions
without the ability of tampering the transaction or history.

\section{Problems with Decentralization}
\seclabel{problems}

We argue that blockchain decentralization introduce several inherent problems.
These problems do not occur only in blockchain providers (physical nodes), 
which much attention has been paid to, but in the full stack.  
\figref{layers} summarizes decentralizations 
in different layers of public blockchains:

\heading{Physical Nodes.} 
The physical nodes, which are in the bottom layer,
consist of a P2P network of miners.
Physical nodes in a public blockchain like Bitcoin and Ethereum are 
assumed decentralized. However,
this assumption does not entirely hold
since ASIC miners and mining pools skew the mining power distribution (\secref{problem1}).

\heading{Platform Software.} 
The platform software layer runs on top of physical nodes.
It includes the implementation of the consensus algorithm 
and the smart contract runtime.
Public blockchain's platform software  
is often developed, maintained, and open sourced
by a community of contributors. 
The impact is in many folds: 
First, the governing and decision making  
is usually led by a ``core'' development community, 
leaders of which
have significant influences on the platform software development. 
Second, 
since the trust of the platform software is critical and economically 
rewarded, 
the development community, in many cases, 
has strong incentives to make the platform reliable.
Third, the platform software layer needs
to be supported by the mining network. 
For example, a new update of the platform software will not happen or
will fork the blockchain
if it is rejected by the mining network owners. 

\heading{Smart Contract.}
Smart contracts or DApps \cite{dappradar} are the applications 
deployed on the blockchain platforms. 
The application logics are encoded in smart contract code.  
The execution of smart contracts is triggered 
by function calls from clients or other smart contracts.
The result of execution is reflected in the state changes of blockchains
(e.g., a change in user's account balance). As to be discussed in \secref{problem3}, 
the execution of smart contract is fully replicated and sequential. 

\heading{Client.}
The client layer serves the end users of smart contracts. 
It takes user inputs and displays the end results to users.
The client layer only needs to interact with the platform layer using standard APIs. 
It is completely open 
since anyone is free to implement their own client.  

\begin{figure}
\centering
\includegraphics[width=0.7\columnwidth]{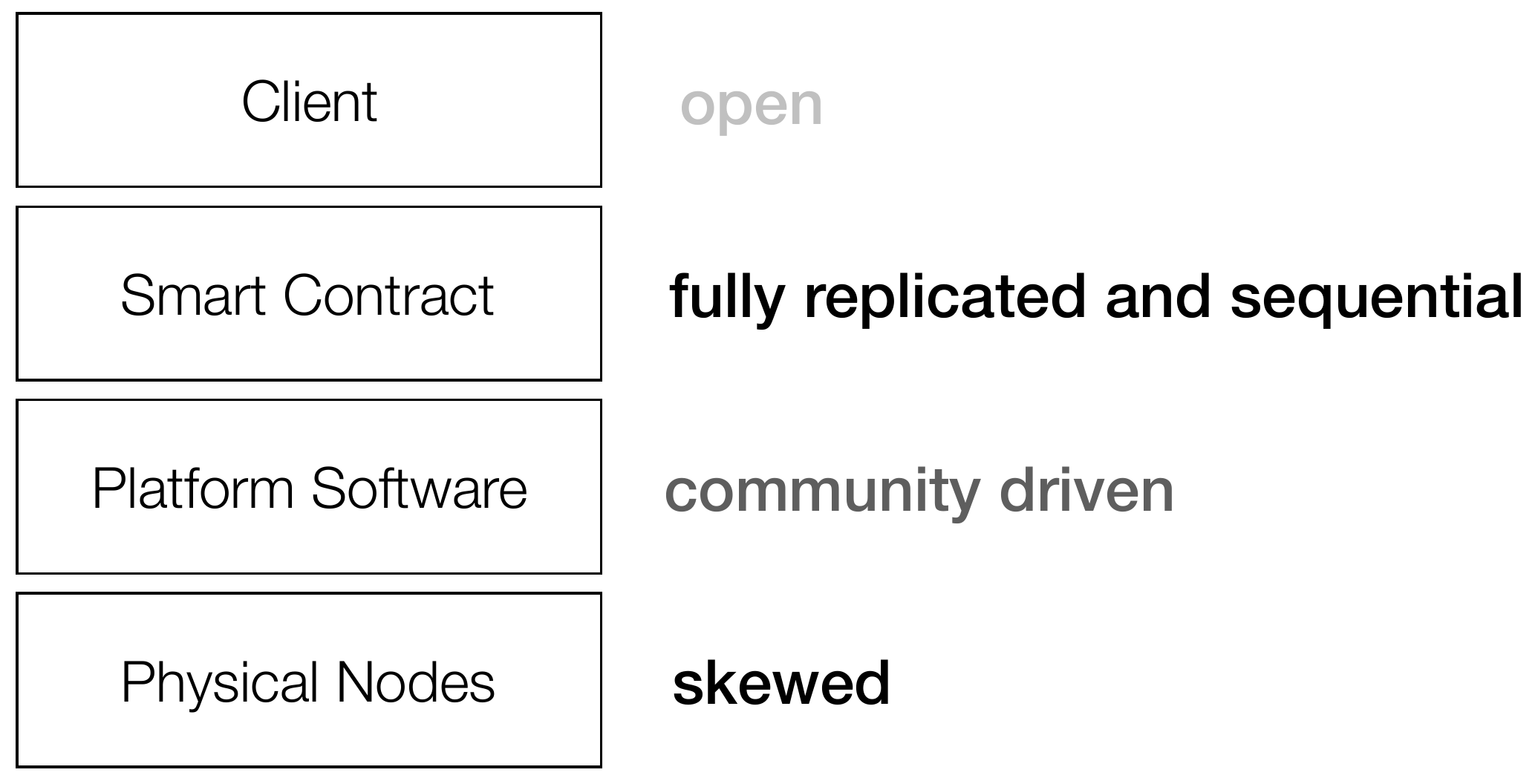}
\caption{Decentralization of full-stack blockchains.}
\figlabel{layers}
\vspace{-0.2in}
\end{figure}

In the rest of this section,
we discuss three major problems with blockchain decentralization,
one in each layer.

\subsection{Skewed Mining Power}
\seclabel{problem1}
The first problem brought by blockchain decentralization is skewed mining power 
in PoW mining network. This problem lays in the physical nodes layer, 
which stays in the bottom of the blockchain stack. 
To leverage the decentralization of the P2P
mining network 
to reach consensus,
a key assumption of Nakamoto consensus \cite {bitcoin} is that each mining node 
has similar computation power 
thus similar probability to extend the blockchain. 
However, there are two trends in mining networks of 
the major blockchain systems: 
First, incentivized by the surging price of cryptocurrencies, 
the mining power of a single mining hardware grows exponentially, 
especially since the introduction of ASICs that are purely designed 
for performing crypto hashing. 
Second, since the number of mining nodes increases dramatically, 
although the expected profit of mining is still high ( 
thanks to the high price of cryptocurrencies) , 
the variance of mining profit increase significantly%
\footnote{One analogy is a lottery with positive expected return. Even if the return is positive, for an individual, it is still a high probability to lose money (invest on mining hardware and electricity but get zero block reward).}.
Thus, miners form mining pools in order to \emph{stabilize} profit.  

As a result, the distribution of mining power is \emph{highly skewed} in real world.
For example, as shown in a recent measuring study \cite{GencerBERG18},
$90\%$ of the mining power is controlled by $16$ miners in Bitcoin ($16_{0.1}$ decentralized) and
$11$ miners of Ethereum ($11_{0.1} decentralized$). Moreover, top $4$ Bitcoin miners have more
than $53\%$ of the mining power in total ($4_{0.47}$ decentralized), 
and top $3$ Ethereum miners have 
more than $61\%$ of the mining power in total ($3_{0.39}$ decentralized). 


This means the blockchain is effectively maintained 
by very few distinct entities. 
Although happens rarely, $51\%$ attack does occur in real world. 
One example is the recent $51\%$ attack \cite{btgattack} to Bitcoin Gold \cite{btg} which leads to $18$ million us dollar worth loss.
The skewed mining power demonstrate 
considerable vulnerability of current public chains. 

\subsection{Scalability of Transaction Throughput}
\seclabel{problem2}
Second, we argue that the decentralized consensus algorithms cause
\emph{inherent scalability} of blockchain transaction throughput, regardless of detailed 
protocol implementations. 
Scalability of blockchain transactions has been witnessed in practice.
For example, the peak transaction throughput of Bitcoin and Ethereum are $3$ txn/sec and 
$15$ txn/sec \cite{etherpeak}, which are insufficient 
for many performance critical applications.
Here, for the first time, we show that this scalability bottleneck is unavoidable, as long as 
the consensus algorithm requires consensus from all participants, e.g. PoW or PoS%
\footnote{DPoS sacrifices decentralization, and thus can achieve better scalability.}. 
To demonstrate that, we prove upper bounds of transaction
throughput of distributed consensus algorithms and 
show that this upper bound is very low in real world settings.  

To derive the upper bounds bound formally, 
we first review the essential definitions of terms in 
blockchain systems.
In public blockchains, 
every participant shares a single global state 
and reaches an agreement (with high probability) on
any computations on the global state. 
As a result, the latency of a transaction, $\mathcal{L}$, is determined 
\footnote{Given the transaction is successfully written to the blockchain.} 
by the number of confirmations required (usually $6$ in practice), $C$, and 
the time interval of block generation, $p$:
\[ \mathcal{L} = C \times p\]
For example, $p_{\text{Bitcoin}} = 10 \text{ mins}$, which leads to $\mathcal{L}_{\text{BitCoin}} = 60 \text{ mins}$;
$p_{\text{Ethereum}} = 15 \text{ secs}$ ($10 \sim 20$ secs in practice), 
which leads to $\mathcal{L}_{\text{Ethereum}} = 1.5 \text{ mins}$.

Similarly, the max transaction throughput $\mathcal{R}$, is determined 
by the time interval of block generation $p$ and the number of transactions in a block $N$.
And $N = b/s$, where $b$ is the block size and $s$ is size of each transaction:
\begin{equation}
 \mathcal{R} = \frac{b}{s \times p} \label{eq:r1} 
\end{equation}
 
For example, the average transaction size 
on the Bitcoin blockchain is $513.86$ bytes 
over the last $6$ years \cite{btctxnsize}, and 
the Bitcoin block size is $1$ MB. 
It is easy to calculate that $R_{Bitcoin} \approx 3.4$ txn/sec.
Similarly, we can get that $R_{ethereum} \approx 15 $ txn/sec, 
which is close to 
Ethereum's peak performance observed (~15 txn/sec, Jan. 4, 2018) \cite{etherpeak}.

Below, we give an upper bound on the ideal transaction throughput that a decentralized 
blockchain can achieve.

\begin{theorem}
Even ignoring the local computation time, the transaction throughput a decentralized 
blockchain system can achieve is less than:
\[ R \leq \frac{w}{s}\]
where $s$ is the size of a transaction on blockchain,
$w$ is the access bandwidth.  
\end{theorem}

\begin{proof}
Let's take a closer look at Eq.~(\ref{eq:r1}). $p$ cannot be infinitely small 
since it takes time to broadcast a block to each node. Thus:
\[ p \geq l + \frac{b}{w}\]
where $l$ is the network latency. 
Using the back of the envelope calculation, we have:
\[ R \leq \frac{b}{s(l+\frac{b}{w})} \]
The only thing that we can adjust here is $b$. Let's look at the derivative on $b$: 
\[ \frac{dR}{db} = \frac{lw^2}{s(lw+b)^2} \]

The derivative is always positive. Thus, $R$ is monotonically increasing with  $b$.
The upper bound of the transaction throughput is when $b \to \infty$:

\[ \lim_{x \to \infty}\frac{x}{s(l+\frac{x}{w})} = \frac{w}{s} \] 
\end{proof} 

It is worth noting that $w$ should be reasonably and conservatively chosen so that
the system allows the majority of the mining nodes to be able to
collect mining rewards. 
In a recent measurement study \cite{GencerBERG18}, 
$67\%$ Bitcoin mining nodes have larger than $23.3$ Mbps access bandwidth; $90\%$ Bitcoin mining nodes have larger than $5.7$ Mbps access bandwidth; the access bandwidth for $67\%$ and $90\%$ 
Ethereum nodes are $11.2$ Mbps and $3.4$ Mbps, respectively. 
So even if we ignore network congestion control,
block verification time, and assuming infinite block size, 
the Bitcoin throughput is at most $1.1$K txn/sec and 
Ethereum throughput is at most $700$ txn/sec (using the $90\%$ users' bandwidth). 
In fact, the actual throughput is much smaller
than this theoretical limit.

\subsection{Scalability of Smart Contract Execution}
\seclabel{problem3}

\begin{figure}
\begin{lstlisting}[language=solidity]
contract TokenContractFragment {
// Balances for each account
mapping(address=>uint256) balances;
// Get the token balance for account `tokenOwner`x
function balanceOf(address tokenOwner) 
         public constant returns (uint balance) {
  return balances[tokenOwner];
}
// Transfer the balance from owner's account 
// to another account
function transfer(address to, uint tokens) 
         public returns (bool success) {
   ...
\end{lstlisting}
\vspace{-0.2in}
\caption{An Example ERC20 Token Contract~\cite{erc20}.}
\figlabel{erc20}
\vspace{-0.2in}
\end{figure}

Apart from the scalability of transaction throughput, smart contract execution 
in current decentralized blockchain systems does not scale as well. 
In particular, the following two problems prevent scaling smart contract execution. 

\heading{Fully Replicated and Single Threaded Execution}
In current decentralized blockchain systems like Bitcoin or Ethereum, 
the effective execution of smart contracts 
(the execution result that is eventually included in the blockchain) is repeated in every 
mining node. So the effective computation power of 
the entire blockchain system is essentially the same as single node. 
In addition, the runtimes of smart contracts (VMs) are single threaded. 
Thus, it is impossible to leverage parallelism within a single node as well.
As a result, the computation power of the entire Bitcoin or Ethereum network, 
which consists of more than hundreds of thousands machines, is less than a modern mobile phone. 

\heading{Sequential Programming Model}
Even if the smart contract execution is parallelized, we argue that it is still hard 
to scale smart contract execution since the current smart contracts are written using a 
sequential programming model. 
For example, \figref{erc20} shows the ERC20 token contract \cite{erc20}, one of the most popular 
smart contracts used in Ethereum for many ICOs.
It is written in Solidity \cite{solidity}, a JavaScript-style language.
It has global variables,  such as \texttt{balances}, 
which stores the token balance of each account.
Given the generality of the smart contract language (Solidity is Turing complete), 
it is challenging to scale up the execution of smart contracts 
as they are currently written.

\section{Research Opportunities}
\seclabel{beyond}

We propose three research directions to solve or circumvent problems with decentralization.

\begin{itemize}
	\item Mining power should be more decentralized by design.
        \item Other forms of trust should be considered, when possible, to replace decentralization.
	\item Smart contract should be scaled to achieve a higher scalability in overall.
\end{itemize}

\subsection{Democratizing Mining Power}

One key factor that leads to skewed mining power is the emergence of specialized 
mining hardware, especially ASICs. 
Specialized mining hardware has outperformed personal computers by orders of magnitude
in terms of both mining power (number of hashes per second) and mining energy efficiency (number of hashes per watt).
In addition, specialized mining hardware is usually very expensive.
As a result, they mostly end up in the hands of a small number of groups,
such as owners of big mining farms and ASIC miner manufacturers.    

One possible approach to democratizing mining power is 
to design ASIC proof hashing algorithms for 
PoW. For example, Litecoin \cite{litecoin} and Dogecoin \cite{dogecoin} use scrypt, 
a hash algorithm that is designed to 
be ASIC proof by its high memory consumption.
However, this proved to be extremely difficult since the ASIC improved as well. 
As an example, in 2014, specialized ASIC mining hardware was launched for scrypt-based
cryptocurrencies \cite{scryptasic}.

\subsection{Trust, Not Necessary Decentralization}

As shown in \secref{problem2}, there is an inherent trade-off 
between decentralization and transaction throughput. 
Ultimately,
decentralization is a means to democratize trust rather than
the goal. 
There are other ways of ensuring trust.
For example,
\emph{verifiable computation} and \emph{secured hardware} 
\cite{walfishVC, sgx}
allows clients to run code on untrusted platforms.
In doing so, an untrusted platform generates verifiable proofs
for clients to vet the correctness of computation.
With verifiable computation, trusting the majority of
system nodes does not need to be assumed any more.

As another example,
\emph{formal verification} and \emph{certified programming}
\cite{sel4, compcert}
allow programmers to provide a mathematical proof showing that
a blockchain implementation meets its specification 
by construction. This could be used to eliminate the unintended behaviors 
in blockchain implementations. 

\subsection{Scaling Smart Contract Execution}
To achieve scalable execution of smart contracts,
it is time to rethink the design of both
the programming model and the runtime.
First, new programming primitives need to be introduced to make parallel execution of 
smart contracts possible.
For example, many programming constructs, such as
concurrent data structures \cite{MoirS04}, can be borrowed 
from extensive programming languages research in past decades. 
Second, the smart contract runtime needs to be redesigned to support parallel execution of smart contracts 
and at the same time still maintains a deterministic transaction order and 
 keeps all the transactional guarantees. 
Many existing approaches in databases and systems should be revisited and adopted 
to smart contract runtime \cite{BernsteinBook}.

\section{Conclusion}

In this paper,
we defined a quantitative measure of blockchain decentralization and comprehensively compared the decentralization aspect of current blockchain systems using this measure. 
We identified key challenges in blockchain decentralization: skewed mining power distribution, 
inherent conflicts between decentralization and scalability (we proved the first upper bound of
transaction throughput of decentralized blockchain systems), and
fully replicated and single threaded execution and sequential programming model prevent scaling smart contract execution.
Finally, we proposed three possible research directions
towards solving these challenges. 
We believe that adopting ideas from many other fields, such as formal methods, 
programming languages/compilers, and databases, 
could provide new approaches to address 
the trade-offs between scalability and decentralization.


\begin{acks}
The authors would like to thank Dr. Xi Wang for his detailed and constructive comments.
\end{acks}

\bibliographystyle{ACM-Reference-Format}
\bibliography{paper}

\end{document}